\documentclass[acmsmall,screen]{acmart}
\AtBeginDocument{%
  }

\usepackage{adjustbox}
\usepackage[table]{xcolor}
\usepackage{tcolorbox}
\usepackage{multirow}
\tcbuselibrary{breakable}
\tcbuselibrary{skins}

\newcommand{\answer}[2]{
  \begin{tcolorbox}[
    enhanced, 
    breakable,
    left=3mm, right=3mm,
    colback=blue!5,
    colframe=blue!60,
    boxrule=0.5pt,
    arc=2pt,
    fonttitle=\bfseries,
    title=Answer for RQ#1:
    ]
    #2
  \end{tcolorbox}
}

\usepackage[normalem]{ulem}

\newcommand{\ours}{\textsc{REprompt}}

\begin{document}

\title{\ours{}: Prompt Generation for Intelligent Software Development Guided by Requirements Engineering}

\author{Junjie Shi}
\email{CHUNKIT001@e.ntu.edu.sg}
\affiliation{%
  \institution{Nanyang Technological University}
  \country{Singapore}
}

\author{Weisong Sun}
\authornote{Corresponding author.}
\email{weisong.sun@ntu.edu.sg}
\affiliation{%
  \institution{Nanyang Technological University}
  \country{Singapore}
}

\author{Zhenpeng Chen}
\affiliation{%
  \institution{Nanyang Technological University}
  \country{Singapore}
}

\author{Zhujun Wu}
\affiliation{%
  \institution{East China Normal University}
  \city{Shanghai}
  \country{China}
}

\author{Xiaohong Chen}
\affiliation{%
  \institution{East China Normal University}
  \city{Shanghai}
  \country{China}
}

\author{Zhi Jin}
\affiliation{%
  \institution{Wuhan University}
  \city{Wuhan}
  \state{Hubei}
  \country{China}}

\author{Yang Liu}
\affiliation{%
  \institution{Nanyang Technological University}
  \country{Singapore}
}

 \renewcommand{\shortauthors}{Shi et al.}

\begin{abstract}
The development of large language models (LLMs) is transforming software development.
LLMs can not only work as code auto-completion tools in an integrated development environment, but also as a foundation model in a coding agent in a vibe-coding scenario.
Consequently, prompts become increasingly crucial in driving agent-based intelligent software development as it not only work as displinary of LLM it self, but holder of user requirement. 
In the dominant conversational paradigm of LLMs~\cite{openai_chatml_github,deepseek_api_docs_2026,anthropic_claude_api_2026}, prompts are divided into system prompt and user prompt, where the system prompt provides high-level instructions to steer the AI's behavior and set the context for the entire conversation, and the user prompt represents the input or questions from the human user.

However, designing effective prompts remains a highly challenging task, as it requires a deep understanding of both prompt engineering and software engineering, particularly requirements engineering. To reduce the difficulty of prompt construction, a series of automated prompt engineering techniques have been proposed. Nevertheless, these approaches largely neglect the methodological principles of requirements engineering, making them inadequate for generating artifacts that conform to formal requirement specifications in software development scenarios.

To address these challenges, we propose \ours{}, a multi-agent prompt optimization framework guided by requirements engineering. Inspired by requirements engineering, we treat prompts from user as initial requirements and employ multiple agents to simulate crucial activities in requirements development. Specifically, requirements development consists of four stages: elicitation, analysis, specification, and validation. Correspondingly, \ours{} is designed with four stages. 

To validate the effectiveness of \ours{}, we conduct experiments on two tasks: system prompt optimization and user prompt optimization. For the system prompt experiment, we examine whether \ours{} can optimize different system prompts. Evaluation is performed using both LLM-as-a-judge and human assessment. 
In this scenario, \ours{} achieves positive improvements across all agents. In LLM-as-a-judge scoring, the optimized role prompts receive scores of up to 4.7 in consistency and 4.5 in communication (out of 5). In human evaluation, we obtain scores of 5.75 in overall user satisfaction and 5.42 in usability (out of 7). These results show that \ours{} effectively improves the usability of role prompts.
For the user prompt experiment, we use YouWare~\cite{youware}, a vibe-coding platform hosting hundreds of thousands of projects, as our test platform. In this scenario, our framework achieves overall satisfaction scores of 6.3 (games subset) and 6.5 (tools subset). Additionally, for the games subset, it achieves 6.17 in usability, 6.4 in information quality, and 6.53 in interface quality (all out of 7), which indicates that \ours{} comprehensively optimizes user prompts.

\end{abstract}

\maketitle

\section{Introduction}
\label{Introduction}

Large language models (LLMs) and LLM-based agents have profoundly transformed software development. Developers can now utilize tools like Copilot~\cite{wermelinger2023using} as integrated development environment plugins to accelerate coding or input their requirements into LLM-based coding agents in vibe-coding scenarios~\cite{ray2025review} to achieve end-to-end generation of software projects. Prompts play a pivotal role in intelligent software development, as they not only define the role, persona, and behavioral guidelines for LLM and coding agents but also serve as carriers of user requirements. In the dominant conversational paradigm of LLMs~\cite{openai_chatml_github, deepseek_api, anthropic_claude_api_2026, qwen_api}, prompts are structured as role-specific messages, where the system prompt provides high-level instructions to steer the AI’s behavior and set the
context for the entire conversation, and the user prompt represents the input or questions from the human user. 

However, crafting effective prompts for software development is challenging because of the misalignment between naive prompt caused by human intent and fine-grained requirements needed for software development~\cite{jin2025conceptual, ullrich2025requirements}. This indicates the need for dedicated requirements engineering methodologies to refine prompt generation, thereby bridging the gap between users’ vague initial requirements and the well-specified requirement specifications necessary for correct software development. However, existing prompt optimization approaches often naively refine initial prompts along simple dimensions, such as making them more detailed, longer, or safer~\cite{dong2023pace,sinha2024survival,yang2023instoptima}.

Therefore, we propose \ours{}, a requirements engineering~\cite{van2000requirements,huang2025knowledgeguidedmultiagentframeworkautomated}-guided prompt optimization framework that supports the refinement of both system and user prompts in agent-based software development. We posit that prompts in software development can be viewed as a form of software requirements. set up four agents in \ours{}, which are Interviewee, Interviewer, CoTer, and Critic. Inspired by the four stages of requirements engineering, which are elicitation, analysis, specification, and validation, we design corresponding phases in \ours{}. We implement requirements elicitation stage as interviewing process between Interviewer and Interviewee for interviewing is commonly used in requirements elicitation. Then in requirements analysis, Interviewer will transform interview record into a draft of software requirements specification. Specifically, we reformulate the target output of requirements specification from software requirements specification into a chain-of-thought structure: for user prompts, the chain takes the form of a strictly ordered and dependency-aware programming task list, while for system prompts, we use predefined agent prompt templates. Finally, in the validation stage, Critic agent is employed to evaluate the generated chain-of-thought.

To validate effectiveness of \ours{}, we conduct experiments on both role prompt optimization and user prompt optimization. In the system prompt experiment, we select a multi-agent architecture example provided within MetaGPT~\cite{hong2024metagpt} as our test case since MetaGPT is the first open-source meta-programming framework. This architecture accepts direct user requirements and generates the corresponding product requirement documents (PRDs) and system design documents (SDDs). To demonstrate the effectiveness of \ours{}, we independently optimize different agents within this architecture and compare whether \ours{} achieves improved results. We employ both LLM-based evaluation and human evaluation as our assessment benchmarks. For LLM-based evaluation, we use an LLM to score the generated documents. For human evaluation, we treat the generated documents as input to YouWare~\cite{youware}, a platform for vibe coding that has hosted more than 100,000 projects to date, and relies on human assessors to rate the final software artifacts. Our framework demonstrates consistent improvements across all evaluated agents. For documentation scoring, the optimized role prompts achieves scores of up to 4.7 in consistency and 4.5 in communication (on a 5-point scale). In artifact evaluation, the framework receives scores of 5.75 for overall user satisfaction and 5.42 for usability (on a 7-point scale). These results indicate that \ours{} is effective in improving both documentation quality and software artifact usability.
For user prompt experiment, we directly optimize the corresponding prompt for each scenario and use the optimized prompt as the input to YouWare. To evaluate the effectiveness of \ours{} in practice, we further employ human evaluation to assess the final software artifacts. \ours{} achieves overall satisfaction scores of 6.3 on the games subset and 6.5 on the tools subset (also on a 7-point scale). Additionally, within the games subset, \ours{} is scored 6.17 in usability, 6.4 in information quality, and 6.53 in interface quality. These evaluation outcomes demonstrate the strong practical effectiveness of \ours{}.

In summary, this paper makes the following contributions:
\begin{enumerate}

\item We incorporate Requirements engineering with LLM-based software development, creating a requirements engineering-optimized prompt framework named \ours{} that supports the entire process. 

\item We do extensive experiment to validate effectiveness of \ours{}. Results show the effectiveness of \ours{} on both user prompt and role prompt.
\end{enumerate}

\section{Background and Related Work}
\label{Related}

\subsection{Requirements Engineering}
Requirements Engineering\cite{wiegers2013software} is a core process in software engineering aiming to systematically identify, analyze, document, validate, and manage the requirements of software systems, with objective to ensure that the developed software system accurately meets the actual needs and expectations of stakeholders. The Requirements Engineering process can be divided into four main stages: requirements elicitation, which involves gathering requirements from stakeholders; requirements analysis, where the elicited requirements are further examined and refined to produce a draft of the software requirements specification; requirements specification, which formalizes the analyzed requirements into a structured Software Requirements Specification to enhance human understanding; and finally, requirements validation, where the specified requirements are reviewed and verified. Requirements Engineering provides a systematic methodology for eliciting, analyzing, and refining human initial requirements to meet the comprehensive and structured demands of engineering contexts. 

Researchers have utilized LLMs to automate requirements acquisition across various aspects, employing methods such as interview generation and user story creation. Feasibility studies indicate that LLMs can be effectively applied to automated requirements generation. Görer et al.~\cite{gorer2023generating} demonstrated that LLMs are capable of generating interview questions that approach human-level quality. Similarly, research by Quattrocchi et al.~\cite{quattrocchi2025can} shows that LLMs can produce high-quality user stories and evaluate their quality. During the requirements analysis phase, Jin et al.~\cite{jin2025system} developed a system modeling benchmark to assess the capability of LLMs in modeling systems from natural language requirements. Lutze et al.~\cite{lutze2024generating} evaluated the performance of various LLMs in generating requirements specifications from demand documents for smart devices. Their findings reveal that while LLMs can generate specifications with high accuracy, they struggle with requirements that contain ambiguity or inconsistencies. Furthermore, Wang et al.~\cite{wang2025supporting} explored the use of LLMs in converting received software requirements specifications into FERT, a language used for formal modeling and acceptance of requirements. The results demonstrate that LLMs can significantly reduce the cost and lower the barrier to entry for formal requirements analysis.

\subsection{Auto Prompt Optimization}
Current automatic prompt optimization approaches can be broadly categorized into search-based prompt optimization methods and feedback-based prompt optimization methods.
Search-based prompt optimization methods typically formulate prompt optimization as a search problem over a discrete prompt space. Some approaches~\cite{prasad2023gripsgradientfreeeditbasedinstruction} perform phrase-level search by applying deletion, reordering, and simplification operations to human-written instructions. Done et al.~\cite{dong2023pace} propose an approach inspired by the actor–critic reinforcement learning framework, where prompts are treated as policies and iteratively refined using model-generated feedback. Wang et al.~\cite{wang2023promptagentstrategicplanninglanguage} further formulate prompt optimization as a planning problem and employ Monte Carlo tree search~\cite{browne2012survey} to strategically explore the prompt space.

Feedback-based prompt optimization methods focus on the availability and reliability of optimization signals in real-world application scenarios. Unlike approaches that rely on numeric scores or labeled datasets, Lin et al.~\cite{lin2024prompt} investigate prompt optimization using only human preference feedback, enabling optimization without explicit reward functions. Furthermore, Xiang et al. ~\cite{xiang2025self}completely remove the assumption of external supervision by adopting an LLM-as-a-judge paradigm, in which outputs generated by different prompts are compared in pairs to guide prompt selection and optimization under a self-supervised setting.

In addition, several empirical studies~\cite{do2025makes,santana2025prompting} systematically evaluate automated prompt engineering techniques across a variety of tasks. These studies suggest that optimizing prompts toward a single objective may be more effective than optimizing for multiple objectives simultaneously. However, none of these works consider prompt optimization in the context of software development tasks.

\section{Design of \ours{}}
\begin{figure}[tbp!]
    \centering
    \includegraphics[width=0.8\linewidth]{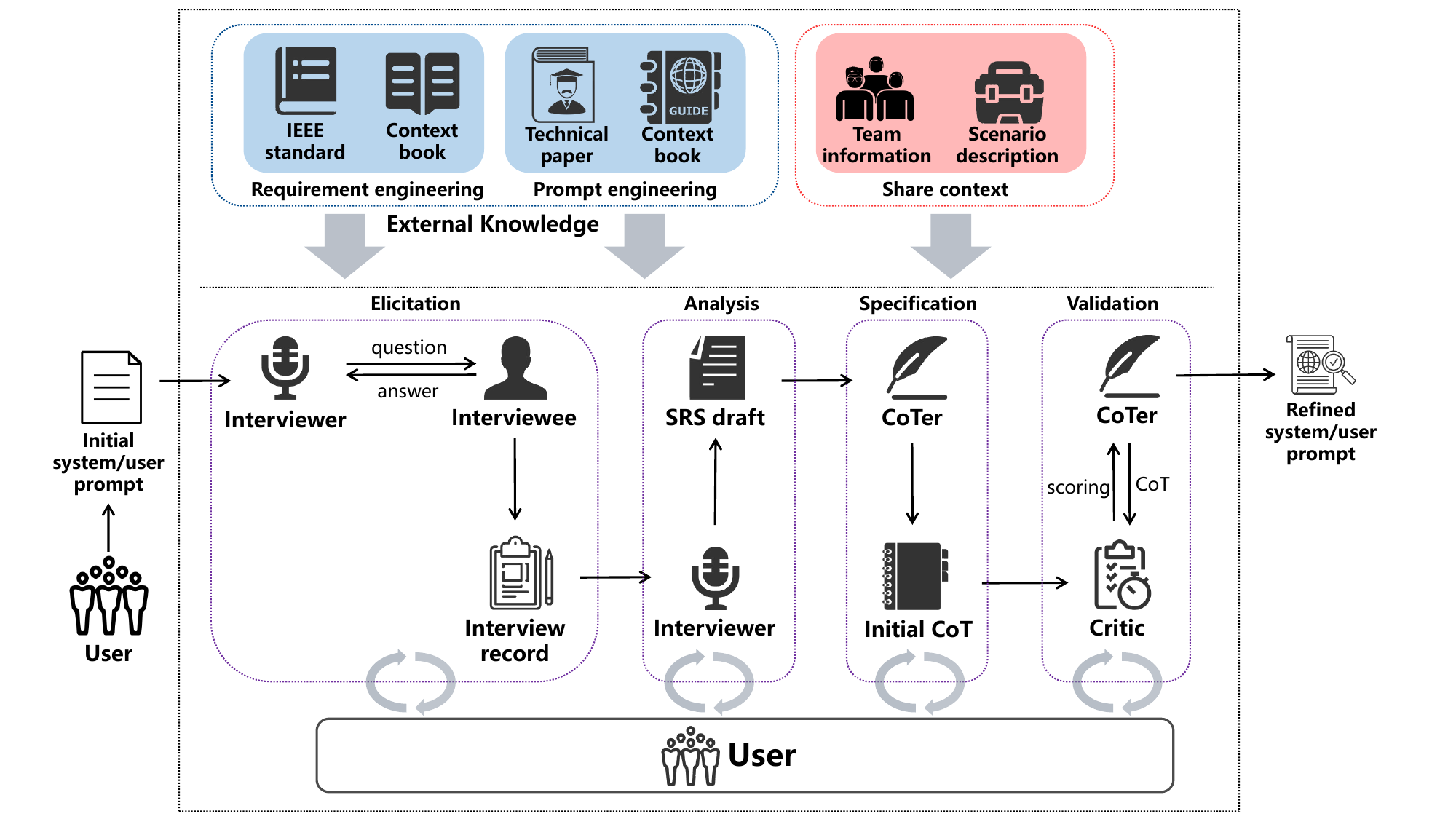} 
    \caption{Workflow of \ours{}.}
    \label{fig:REprompt}
\end{figure}

As illustrated in \autoref{fig:REprompt}, \ours{} takes either system prompt or user prompt as initial input and output a more comprehensive system or user prompt as output. \ours{} can be divided into requirements elicitation, requirements analysis, requirements specification, requirements validation.  Correspondingly, \ours employs four agents to simulate the requirements engineering workflow: an \textbf{Interviewee agent} that supplements potential user needs from the user’s perspective; an \textbf{Interviewer agent} that elicits system requirements through structured questioning; a \textbf{CoTer agent} that transforms  interview record into structured Chains-of-Thought; and a \textbf{Critic agent} responsible for reviewing and refining the CoT. Further, we incorporate a human-in-the-loop~\cite{wu2022survey} mechanism at the end of each stage to introduce human feedback when hallucinations occur in large language models. At every stage, user confirmation is required before proceeding to the next stage; otherwise, the current stage is re-executed.

\label{framework}

For the requirements elicitation phase, given that conducting interviews with stakeholders is a common practice in requirements engineering~\cite{goguen1993techniques}, we design an interactive process between the Interviewee and Interviewer agents to comprehensively gather user requirements. This phase ultimately produces an interview record. During the requirements analysis phase, after the interview process concludes, the Interviewer agent reviews interview record and produces a draft of the software specification. In the requirements specification phase, we note that traditional requirements engineering in this phase focuses more on adjusting the document format to improve readability for different human stakeholders, rather than modifying the content of the requirements,  therefore, instead of requiring the output to be in the form of a formal software requirements specification, we require it to be formatted according to our designed prompt templates. 

\begin{figure}[tbp!]
    \centering
    \includegraphics[width=\linewidth]{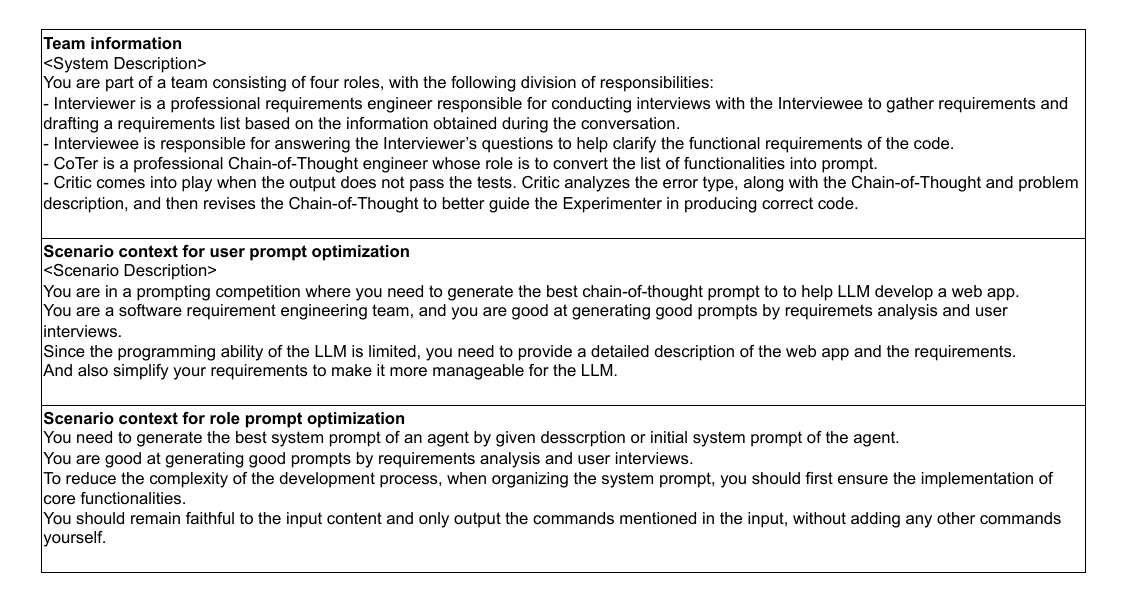}
    \caption{Global shared context of \ours{}.}
    \label{fig:Team_info}
\end{figure}

In addition, \ours{} draws knowledge from existing IEEE 29148-2018~\cite{januaritaiso} standard to acquire comprehensive knowledge of requirements engineering. This standard specifies key processes such as requirements elicitation, analysis, specification, verification, validation, and management, providing a standardized framework and procedures for requirements engineering activities in system and software engineering. Regarding knowledge source of prompt engineering, our references include previous work~\cite{ullrich2025requirements} and expert prompt engineering manuals~\cite{googlewhite} to provide comprehensive understanding of prompt engineering. Additionally, we have designed globally shared context, which an introduction to our agent system team and descriptions of various task scenarios, throughout the entire system to help the agent system better comprehend the context of the tasks it receives. Our system prompts are illustrated in \autoref{fig:Team_info}.

\subsection{Requirements Elicitation}
In the requirements elicitation phase, we obtain requirements through interviews conducted between an Interviewer agent and an Interviewee agent. The Interviewer agent accepts the user’s initial requirements and, following our designed interview process, generates additional potentially relevant questions to pose to the Interviewee. To better organize system development requirements and thereby improve their systematicity and clarity, we draw inspiration from Model-Based Systems Engineering~\cite{wymore2018model} (MBSE). Specifically, the interview process first establishes definitions of software system components and subsequently constructs a requirements model of the system. Since MBSE has been widely applied to the modeling and analysis of complex systems, we argue that it is also well-suited for software development scenarios. The job of the Interviewee agent is to respond to the Interviewer agent’s questions, acting as a simulator of user requirements. Our intuition is that, in traditional requirements engineering, the structuring of raw requirements is typically carried out during the requirements analysis phase by professional engineers, as the elicitation process is constrained by human effort and cost. However, this limitation does not apply to LLMs. As a result, requirements can be partially structured already during the elicitation phase. Therefore, when the Interviewee agent provides responses, we require that these responses be expressed in a partially structured form, making the raw requirements easier for the Interviewer agent to analyze.

\paragraph{Interview Questioning.} Specifically, we designed a four-step interview process. In the first step, we expect the interviewer to inquire about the basic components that may be involved in the software, how each component functions, and how the components should interact with each other. In the second step, we expect the interviewer to ask about the core functionalities that the application should provide based on requirements, the workflow of the application, the minimal set of functions required to achieve these core functionalities, the workflow of these functions, and how they should interact with each other. In the third step, the interviewer will inquire about additional features beyond the core functionalities that could enhance user experience and software quality, the relationship between these features and the core functionalities, and how these features should be implemented. Finally, to prevent uncontrolled expansion of the software scope, we also expect the interviewer to determine which features should be implemented immediately and which can be deferred for future implementation during the third step. In the fourth step, the interviewer will ask about potential front-end requirements to determine the style and presentation methods of the front-end. Detailed information that needed to be confirmed regarding the front-end includes page layout, typography, and the specifics of chart shapes, among others. Lastly, we included an optional confirmation item, namely user guidance, to enhance user control while managing the duration of the interview process. For each step of the interview, we provided corresponding sample questions and explanations of their purposes.

\begin{figure}[tbp!]
    \centering
    \includegraphics[width=0.8\linewidth]{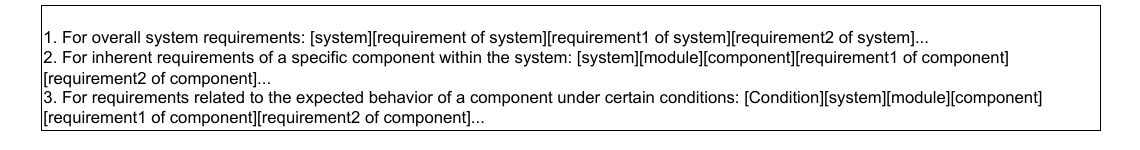} 
    \caption{Requirements template for interviewee agent.}
    \label{fig:REtemplate}
\end{figure}

\paragraph{Interview Answering.} Specifically, drawing on the IEEE 29148~\cite{januaritaiso} standard, we instruct the agent to use objective and precise language when representing user requirements, minimizing the use of comparatives, adjectives, and vague references. Additionally, to bridge the descriptive gap between user expressions, system design, and prompt engineering, we have designed three distinct types of requirements templates to address three different scenarios: overall system requirements, constant requirements for system components, and requirements for system components under specific conditions. The templates are illustrated in \autoref{fig:REtemplate}.
\subsection{requirements Analysis}
In requirements analysis phase, the Interviewer is responsible for transforming the collected requirements into a preliminary draft of the Software Requirements Specification (SRS). To generate a more structured and engineering-compliant SRS, domain knowledge regarding software requirements specifications is incorporated from the IEEE 29148 standard. This knowledge pertains to the organizational methods, methodologies, and perspectives for constructing an SRS, as well as the formal entity categories and their interrelationships that should be included. Additionally, a template for software requirements specifications from IEEE 29148 is adopted as an example framework to guide the Interviewer in generating the SRS draft.

\subsection{requirements Specification}
In requirements specification, CoTer takes the SRS draft and transforms it into a initial version of chain-of-thought.  CoTer operates in two distinct modes, corresponding to two different scenarios: user prompt generation for software development and role prompt generation for agents.

\paragraph{User prompts for software development.} As an empirical study~\cite{ullrich2025requirements} on using LLMs for software development demonstrates, decomposing programming tasks from direct user inputs can significantly enhance the performance of LLMs on programming tasks. Therefore, we employ a programming task list as the format of output in this phase. Specifically, we use a JSON-formatted task list to represent the chain-of-thought for software engineering. To better coordinate contextual dependencies, we expect CoTer to prioritize programming tasks according to software dependency relationships to avoid dependency errors. We stipulate that, prior to formally executing the software writing process, the development of software-related requirements documents and environment setup should be completed first. Additionally, we specify that the entry file of the entire program should be implemented as the final component.

\paragraph{System prompt.} Another scenario is system prompt. We decompose an agent’s system prompt into the following five components:  
\begin{itemize}
\item Role Definition: Defines the responsibilities and tasks of the role.  
\item Knowledge: Essential domain information that the agent must understand and utilize to accomplish a given domain-specific task.  
\item Specification of Available Tools: Describes the tools that the agent can use.  
\item Context information: The context that the agent works in, including work scenario and team information (if agent is part of a multi-agent system).
\item Overview of Work Modes: An agent may have multiple work modes; each mode should be described in sufficient detail. Each mode overview should include following sub-elements: behavior code of conduct,  and examples to help the agent understand the task more thoroughly.  
\end{itemize}
To effectively organize external input into these five components, we require the CoTer to generate the system prompt according to the following process: First, analyze the role that the agent is intended to play in the original prompt and determine the necessary capabilities and workflow for that role, forming a preliminary system prompt.  After rigorously verifying that this preliminary prompt aligns with the given task and original input, incorporate feedback from the Critic and refer to the Software Requirements Specification to organize the content into the five provided components. Each component is then optimized individually.  After performing a completeness check on each section, the CoT is generated according to a predefined template.

\subsection{Requirements Validation}
In requirements validation phase, Critic agent improves the chain-of-thought generated by CoTer by structural scoring. Inspired by wieger et al.~\cite{wiegers2013software}, we let Critic agent to consider the following four aspects to check if Critic agent satisfies RequirementS engineering principle: Completeness, Correctness, Organization and Traceability, Quality Attributes. On the other way, we also let Critic agent to consider the following principles from prompt engineering guidelines~\cite{best_practice_openai}: clear, concise, and consistency.
Upon receiving feedback from the Critic, CoTer also incorporates the Critic’s comments as references to generate improved prompts again.

\paragraph{Structural Scoring.} To achieve clearer and less ambiguous expression, drawing on practices that have been widely used in previous work~\cite{do2024prompt,dong2023pace,sinha2024survival}, we require the Critic to perform structured scoring according to specified steps and provide justifications. The scoring process can be divided into two steps: First, we ask the Critic to evaluate aforementioned aspects of our prompt, including clarity of expression and degree of structure. To meet the requirements of software engineering scenarios, we additionally request the Critic to assess the level of technical detail and executability. At the end of the first step, we require the Critic to summarize the accepted chain of thought, outlining its strengths and weaknesses. Second, we ask the Critic to score each part of the chain of thought based on the previous step's evaluations and finally output the results according to a structured template we provide.
\section{Experiment Setting}
\begin{table}[tbp]
\centering
\caption{APPDev Test Set: Application Scenarios and Corresponding Prompts}
\label{tab:appdev}
\scalebox{0.8}{%
\begin{tabular}{c|ll}
\toprule
\textbf{ID}  & \textbf{Application Scenario} & \textbf{User Prompt}\\
\midrule
1 & go game & I want a go game  \\
2 & chess game& I want a chess game  \\
3 & 2048 game & I want a 2048 game  \\
\rowcolor{gray!30}
4 & tetris game & I want a tetris game  \\
5 & blackjack game & I want a blackjack game  \\
6 & hanoi game & I want a hanoi tower game  \\
7 & snake game & I want a snake game  \\
8 & minesweeper game & I want a mine sweeper game  \\
9 & tic tac toe game & I want a Tic-Tac-Toe game  \\
\rowcolor{gray!30}
10 & sliding block game & I want a sliding block puzzle game  \\
\midrule
11 & clock in software & I want a clock-in app  \\
12 & image format conversion software& I want a Image format conversion tool  \\
\rowcolor{gray!30}
13 & calculator software & I want a calculator software  \\
14 & recording software & I want a recording software  \\
\rowcolor{gray!30}
15 & alarming software & I want an alarm clock software  \\
16 & sentiment analysis software & I want a text sentiment analysis tool  \\
17 & charting software & I want a charting tool  \\
18 & personal website & I want a personal website  \\
19 & word cloud software& I want a word cloud tool  \\
20 & drawing software & I want a drawing app  \\
\bottomrule
\end{tabular}%
}
\end{table}
\label{experiment}
To validate the effectiveness of \ours{} in the domain of software engineering, we propose the following research questions, which will be thoroughly examined in this section:

\paragraph{RQ1: How effectively can \ours{} effectively optimize agent role prompt in multi-agent system in the context of software development?} This question aims to evaluate whether \ours{} can improve agent role prompt in context of multi-agent system for software development task.

\paragraph{RQ2: How effectively can \ours{} optimize user prompts?} Here, we investigate the framework’s capacity to refine direct prompts from users to achieve better performance on software development task.

\paragraph{RQ3: Are all four phases of requirements engineering effective components of the framework?} This research question examines the individual contribution and necessity of each stage, Elicitation,  Analysis, Specification, and Validation, within our automated agent-based approach. We seek to determine whether each phase meaningfully enhances the overall process of prompt generation and optimization.

\paragraph{RQ4: How performance of \ours{} vary with different foundation LLMs?} This examines whether \ours{} can cause improvement on different foundation LLM.

\subsection{Dataset construction}
To evaluate the effectiveness of \ours{} in common scenarios, we collect a set of prompts from 20 scenarios, which can be categorized into two major types: games and tools. Our dataset is called \textbf{APPdev}, the content of \textbf{APPdev} is shown in the \autoref{tab:appdev}. These two categories are selected because game software represents one of the primary revenue sources in APP industry~\cite{wifitalents2025appindustry}, while utility applications are among the most widely installed types of software on user devices~\cite{buildfire2025appstatistics}. Considering the distinction between simple software development and commercial software, we collected ten scenarios from each category with reference to dataset of previous work~\cite{hong2024metagpt,jin2025iredevknowledgedrivenmultiagentframework}.  

In the APPDev game category, the prompts often require the development of an entire game, characterized by high user interactivity and complex game logic. Therefore, we consider the tasks in the game subset to be more challenging. In contrast, the office scenarios primarily focus on needs that may arise in workplace settings, which are relatively simpler in terms of functional complexity and involve lower levels of user interaction.

\subsection{Experiment Design}

For RQ1, we validated the effectiveness of \ours{} on MetaGPT. To the best of our knowledge, MetaGPT is the first open-source framework for multi-agent system programming, and its GitHub repository has garnered 58.3k stars to date. Therefore, we selected it as the platform for constructing the multi-agent system. 

Specifically, we utilize an example multi-agent framework provided by MetaGPT designed for drafting product requirements documents (PRD) and system design documents (SDD). The original prompt for the MetaGPT is manually crafted by experts in prompt engineering. The framework consists of three agents: the first is a team leader, responsible for determining task content based on received input and assigning tasks accordingly. The second is a product manager, who, upon receiving tasks from the team leader, decides whether to conduct market research or directly draft the product requirements documents based on the task requirements. The last is an architect, who writes the SDD based on the tasks assigned by the team leader and the PRD provided by the product manager. We test \ours{} on this multi-agent framework.

We evaluate both the PRDs and SDDs using two methods: human assessment and LLM-as-a-judge~\cite{gu2024survey}. Considering the cost associated with human evaluation, we divide the two software scenarios into two buckets each based on the software size output by YouWare~\cite{youware}, and select one process from each of the four buckets for human evaluation to ensure representativeness in terms of software scale. For all scenarios, LLM-based scoring is applied. To mitigate bias in the LLM evaluations, we utilize G-Eval~\cite{liu2023g} for scoring. For the human evaluation, We recruited eight volunteer participants from East China Normal University to conduct user evaluations of the software. After reviewing a description of the software’s core functionalities, participants were asked to explore the system to assess its functionality, usability, and other related qualities. Each software evaluation session was allocated one hour, and we ensured that each software system was evaluated by at least two participants.

For RQ2, we test \ours{} on YouWare and similarly conduct human evaluations. YouWare is an vibe coding community tailored for individual developers, supporting the direct generation of web applications through natural language. The community currently hosts over 100,000 web projects. Therefore, we consider this platform suitable for examining whether the optimized prompts lead to better software artifacts.

For RQ3, we perform an ablation study by sequentially removing each of the four stages of requirements engineering to observe the impact of different stages on the final output artifacts.

For RQ4, we test \ours{} on different foundation models to confirm whether improve caused by reprompt can be validated on other foundation modelsS.

\subsection{Experiment Metrics.}

\paragraph{LLM as a judge.} Inspired by existing IEEE standards~\cite{5981339} and expert literature~\cite{cagan2005write,or1979universal}, we developed specific evaluation criteria for each type of document. To mitigate bias in the LLM-based evaluation, we employed G-Eval~\cite{liu2023g} for scoring. For the PRD, we use three criteria for evaluation:
\begin{itemize}
     
\item Completeness(Comp.): A comprehensive PRD should include the product’s purpose, features, performance requirements, user needs, and other relevant information.

\item Clarity(Cla.): A well-written PRD should be articulated clearly, with all requirements prioritized in a logical manner.

\item Cohesiveness(Coh.): A high-quality PRD should be cohesive throughout.
\end{itemize}

For the SDD, three corresponding criteria are applied:
\begin{itemize}
 
\item Integrity(Int.): The document should not focus solely on code structure but should provide comprehensive coverage of all aspects of the design.

\item Communicativeness(Comm.): A good SDD should facilitate a shared understanding of the system architecture among stakeholders with diverse backgrounds through structured and clear descriptions.

\item Consistency(Con.): All content included in the SDD must be consistent and free of contradictions.
\end{itemize}

The six metric scores all range from 1 to 5.
\paragraph{Human assessment.} We recruited eight volunteer participants from East China Normal University to conduct user evaluations of the software. In RQ1, human evaluators rate the generated software with Computer System Usability Questionnaire (CSUQ)~\cite{lewis1995computer}, a widely used subjective evaluation instrument for systematically measuring the ease of use and perceived quality of user experience of software systems, information systems, and interactive systems from the user’s perspective~\cite{HajesmaeelGohari2022mHealthQuestionnaires,RussJara2025GuideUsabilityQuestionnairesHealthIT}. The detailed content of CSUQ is shown in \autoref{tab:csuq_items}.

The CSUQ comprises a total of nineteen questions, which can be further grouped into four metrics: 
\begin{itemize}
    \item Overall satisfaction (Ove.): Reflecting the overall satisfaction of the user.
    \item Usability (Use.): Reflecting the functional availability of the software system; 
    \item Information Quality (Info.): Indicating whether the software system provides sufficient guidance for users; 
    \item Interface Quality (Inte.): Reflecting whether the software system offers high-quality front-end pages. We use these four metrics to evaluate the software. 
\end{itemize}
The four metric scores all range from 1 to 7.

Considering the cost challenges associated with large-scale manual scoring, we categorize the two scenarios in APPDev into two buckets, large-scale and small-scale, based on the size of the output projects after direct input into YouWare. From each bucket, one representative sample is select to ensure the representativeness of the chosen datasets. The four selected datasets are presented in gray in \autoref{tab:appdev}.

\subsection{Experiment settings}
We employ Qwen2.5-Max~\cite{qwen2.5-max} as our foundation model, with the temperature parameter set to 0 and the max\_token parameter configured to 4096.

\section{Experiment Result}
\subsection{RQ1: Prompter effectively optimizes agent role prompts.}

\begin{table}[tbp]
\centering
\scriptsize
\caption{Comparison of evaluation metrics on PRD and SDD: \ours{} leads to comprehensive improvements across all evaluation metrics for document scoring. }
\begin{adjustbox}{max width=\textwidth}
\setlength{\tabcolsep}{1pt}
\begin{tabular}{l|ccc|ccc|ccc|ccc|ccc|ccc|ccc}
\toprule
& \multicolumn{12}{c|}{SDD experiment} & \multicolumn{9}{c}{PRD experiment} \\
\midrule
& \multicolumn{3}{c|}{Experiment 1} & \multicolumn{3}{c|}{Experiment 2} & \multicolumn{3}{c|}{Experiment 3} & \multicolumn{3}{c|}{Experiment 4} & \multicolumn{3}{c|}{Experiment 1} & \multicolumn{3}{c|}{Experiment 2} & \multicolumn{3}{c}{Experiment 3} \\
Application Scenario & Int. & Comm. & Con. & Int. & Comm. & Con. & Int. & Comm. & Con. & Int. & Comm. & Con. & Comp. & Cla. & Coh. & Comp. & Cla. & Coh. & Comp. & Cla. & Coh. \\
\midrule
go & 3 & 5 & 5 & 4 & 4 & 5 & 4 & 4 & 5 & 3 & 4 & 4 & 4 & 4 & 4 & 3 & 5 & 4 & 4 & 5 & 5 \\
chess & 3 & 4 & 4 & 4 & 5 & 5 & 4 & 4 & 5 & 4 & 4 & 5 & 4 & 5 & 4 & 4 & 4 & 4 & 4 & 5 & 4 \\
2048 & 4 & 4 & 5 & 3 & 4 & 5 & 4 & 5 & 4 & 4 & 4 & 5 & 4 & 4 & 4 & 3 & 5 & 4 & 5 & 5 & 4 \\
tetris & 5 & 4 & 4 & 3 & 4 & 4 & 4 & 5 & 5 & 3 & 4 & 5 & 4 & 4 & 4 & 5 & 5 & 4 & 4 & 4 & 4 \\
blackjack & 3 & 5 & 4 & 3 & 4 & 5 & 4 & 5 & 4 & 3 & 4 & 4 & 4 & 4 & 4 & 4 & 4 & 4 & 5 & 4 & 4 \\
hanoi & 3 & 4 & 4 & 4 & 5 & 5 & 3 & 4 & 4 & 2 & 4 & 5 & 4 & 4 & 4 & 4 & 4 & 5 & 4 & 4 & 4 \\
snake & 3 & 4 & 4 & 4 & 4 & 5 & 4 & 5 & 5 & 4 & 4 & 5 & 3 & 4 & 4 & 5 & 4 & 5 & 4 & 5 & 4 \\
minesweeper & 4 & 4 & 4 & 3 & 5 & 5 & 3 & 4 & 4 & 4 & 4 & 4 & 4 & 4 & 5 & 4 & 5 & 4 & 4 & 5 & 4 \\
tic tac toe & 3 & 4 & 4 & 3 & 4 & 5 & 4 & 4 & 5 & 3 & 5 & 5 & 4 & 4 & 5 & 4 & 5 & 4 & 5 & 5 & 4 \\
\rowcolor{gray!30}
sliding block & 3 & 4 & 5 & 4 & 4 & 5 & 4 & 5 & 5 & 4 & 4 & 4 & 4 & 4 & 4 & 4 & 5 & 5 & 4 & 4 & 4 \\
clock in & 4 & 5 & 4 & 4 & 4 & 4 & 4 & 5 & 4 & 3 & 5 & 5 & 3 & 4 & 4 & 4 & 4 & 4 & 4 & 4 & 4 \\
image format conversion & 3 & 4 & 4 & 4 & 5 & 5 & 3 & 4 & 4 & 4 & 4 & 4 & 2 & 4 & 4 & 4 & 4 & 4 & 4 & 5 & 4 \\
calculator & 3 & 4 & 5 & 4 & 4 & 5 & 3 & 5 & 4 & 3 & 4 & 5 & 4 & 4 & 4 & 4 & 4 & 4 & 5 & 4 & 5 \\
recording & 3 & 4 & 5 & 3 & 4 & 4 & 4 & 4 & 4 & 4 & 5 & 4 & 4 & 4 & 4 & 4 & 5 & 4 & 4 & 5 & 4 \\
alarming & 3 & 4 & 4 & 4 & 5 & 5 & 4 & 4 & 4 & 4 & 4 & 4 & 4 & 4 & 4 & 4 & 4 & 4 & 4 & 4 & 4 \\
sentiment analysis & 3 & 4 & 5 & 5 & 4 & 5 & 4 & 5 & 4 & 4 & 4 & 4 & 4 & 4 & 5 & 4 & 4 & 5 & 4 & 5 & 4 \\
charting & 4 & 4 & 5 & 3 & 4 & 4 & 4 & 5 & 4 & 4 & 4 & 4 & 4 & 4 & 3 & 4 & 5 & 4 & 4 & 4 & 5 \\
personal website & 4 & 4 & 4 & 4 & 4 & 4 & 4 & 4 & 4 & 4 & 4 & 4 & 4 & 4 & 4 & 4 & 4 & 4 & 4 & 4 & 4 \\
word cloud & 4 & 4 & 5 & 3 & 4 & 5 & 3 & 5 & 4 & 3 & 4 & 5 & 5 & 4 & 4 & 4 & 4 & 4 & 4 & 4 & 4 \\
drawing & 4 & 4 & 4 & 3 & 4 & 4 & 4 & 5 & 5 & 3 & 4 & 5 & 4 & 4 & 4 & 4 & 4 & 4 & 5 & 4 & 4 \\
\midrule
Avg. & 3.45 & 4.15 & 4.40 & 3.60 & 4.25 & \textbf{4.70} & \textbf{3.75} & \textbf{4.55} & 4.35 & 3.53 & 4.16 & 4.53 & 3.85 & 4.05 & 4.10 & 4.0 & 4.40 & \textbf{4.20} & \textbf{4.25} & \textbf{4.45} & 4.15 \\
\bottomrule
\end{tabular}
\label{tab:RQ1SD}
\end{adjustbox}
\end{table}
For RQ1, as mentioned above, we employ a combination of two methods, human evaluation and LLM-as-a-judge, to assess whether the system prompts of the agent have been effectively optimized. The experimental procedure is illustrated in the accompanying figure. For intermediate artifacts generated by \textbf{MetaGPT}, such as product requirements documents and SDDs, we first use the aforementioned LLM-as-a-judge approach to assign scores. Subsequently, representative scenarios are selected, and these documents are used as input for code generation via YouWare. The resulting code is then evaluated through manual scoring.  

\paragraph{Document scoring experiment with LLM-as-a-judge.} Our experimental design comprises a total of four groups:
\begin{itemize}
\item Experiment 1: The original agent system provided by MetaGPT.
\item Experiment 2: Based on the original MetaGPT agent system, the default prompt for the Product Manager agent is replaced with our optimized prompt.
\item Experiment 3: Based on the original MetaGPT agent system, the default prompt for the Team Leader agent is replaced with our optimized prompt.
\item Experiment 4: Based on the original MetaGPT agent system, the default prompt for the Architect agent is replaced with our optimized prompt.
\end{itemize} 
Since the PRD is generated before the Architect agent begins its work, the quality of the PRD is not measured in Experiment 4. Our experimental results are as presented in \autoref{tab:RQ1SD}.

\begin{table}[tbp]
\centering
\scriptsize
\caption{Result for software scoring experiment with human assessment: optimizing both the Team Leader and the Architect led to comprehensive improvements.}
\setlength{\tabcolsep}{3pt}
\label{tab:humanRQ2}
\begin{tabular}{l|cccc|cccc|cccc|cccc}
\hline
\textbf{Application Scenario} & \multicolumn{4}{c|}{\textbf{None}} & \multicolumn{4}{c|}{\textbf{Team Leader}} & \multicolumn{4}{c|}{\textbf{Product Manager}} & \multicolumn{4}{c}{\textbf{Architect}} \\
\cline{2-17}
 & Ove.& Use. & Info. & Inte. & Ove.& Use. & Info. & Inte. & Ove.& Use. & Info. & Inte. & Ove.& Use. & Info. & Inte. \\
\hline
2048 & 5& 5.42 & 6.33 & 5.00 &7& 5.83 & 6.67 & 6.33 & 5& 5.50 & 6.33 & 5.67 & 6 & 5.83 & 6.67 & 6.00 \\
\rowcolor{gray!30}
sliding block & 2 & 3.00 & 2.00 & 2.00&6 & 5.50 & 5.33 & 6.00 & 4& 4.75 & 6.00 & 3.33 & 6 & 5.08 & 6.00 & 5.33 \\
calculator & 6 & 5.67 & 5.00 & 5.33&5 & 4.75 & 3.00 & 5.33 & 4& 3.92 & 2.67 & 4.33 & 6 & 4.83 & 3.67 & 5.67 \\
alarming &5 & 5.75 & 6.00 & 6.00 &5& 5.42 & 5.67 & 5.67 & 6&5.58 & 5.67 & 6.00 & 5 & 5.92 & 6.67 & 6.00 \\
\hline
avg & 4.5 & 4.96 & 4.83 & 4.75 & \textbf{5.75} & 5.38 & 5.17 & \textbf{5.83} & 4.75 & 4.94 & 5.17 & 4.83 & \textbf{5.75} & \textbf{5.42} & \textbf{5.75} & 5.75 \\
\hline
\end{tabular}
\end{table}

\paragraph{Software scoring experiment with human assessment.} As mentioned earlier, we used the artifacts generated by MetaGPT, namely the PRD and SDD, as input to YouWare to evaluate the quality of the documents. To demonstrate the effectiveness of \ours{}, we sequentially replaced the original expert prompts of the three agents in MetaGPT with the optimized agent prompts. Therefore, the numbering of our experimental groups corresponds to that used in the document scoring experiments employing LLM-as-a-judge. Our experimental results are as
presented in \autoref{tab:RQ1SD}.

\paragraph{Experiment result for document scoring experiment with LLM-as-a-judge.} As shown in Table 2, the experimental results of using LLM for document scoring demonstrate that for both PRD and SDD, optimizing any one of the three agents in \ours{} leads to comprehensive improvements across all evaluation metrics for document scoring. In the SDD scoring experiment, we observed that optimizing the Product Manager resulted in the greatest enhancement in the consistency metric of the SDD, increasing from 4.4 to 4.7. Meanwhile, optimizing the Team Leader led to the most significant improvements in integrity and communication, with scores rising from 3.45 to 3.75 and from 4.15 to 4.55, respectively. Similar trends are observed for the software requirements document. Optimizing the Product Manager resulted in the largest gain in the cohesion metric of the SDD, improving from 4.1 to 4.2. In contrast, optimizing the Team Leader yielded the most notable improvements in the completeness and clarity of the product requirements document, with scores increasing from 3.85 to 4.25 and from 4.05 to 4.45, respectively.  

Therefore, we conclude that optimizing the Team Leader primarily enhances the completeness of content and clarity of expression in the documents, while optimizing the Product Manager helps constrain the scope of content, thereby improving consistency and cohesion. Overall, for the document scoring task, although \ours{} comprehensively improves all metrics regardless of which agent in the MetaGPT is enhanced, the Team Leader contributes the most significant overall improvement across both tasks. This may be because, although the Team Leader does not directly draft the documents, it can indirectly enhance document quality by refining the task descriptions assigned to the other agents.

\begin{figure}[tbp!]
    \centering
    \includegraphics[width=0.8\linewidth]{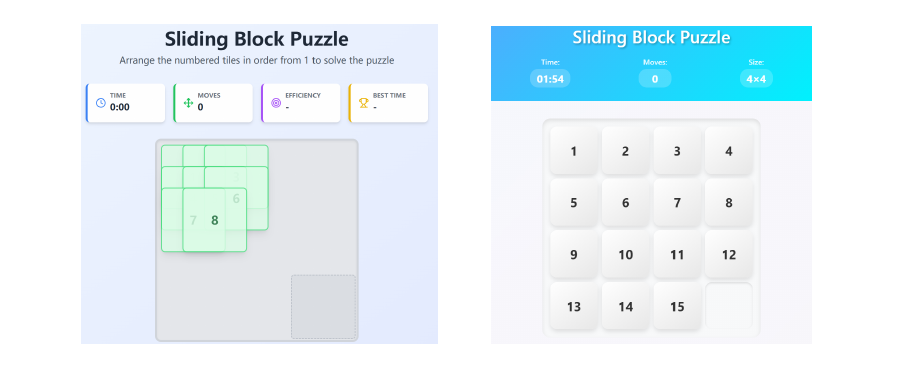} 
    \caption{The interface of the software artifact under the configuration of Experiment 1 and Experiment 2 is presented, with the software artifact of Experiment 1 on the left and that of Experiment 2 on the right.}
    \label{fig:sliding_block_puzzle}
\end{figure}

\paragraph{Experiment result for human assessment experiment with LLM-as-a-judge.}
As shown in \autoref{tab:humanRQ2}, our results indicate that optimizing both the Team Leader and the Architect led to comprehensive improvements across various software product metrics. For instance, both optimizations resulted in a significant increase in overall user satisfaction, from 4.5 to 5.75. Specifically, the optimization of the Team Leader contributed most notably to the enhancement of the interface quality metric, which rose from 4.75 to 5.83. On the other hand, the optimization of the Architect yielded the greatest improvements in usability and information quality metrics, increasing from 4.96 to 5.42 and from 4.83 to 5.75, respectively. However, we observed that optimizing the Product Manager did not lead to significant gains in user satisfaction or usability. We believe that this is caused by the fact that in relatively simple software systems, improvements in consistency and cohesion do not substantially enhance software usability.

\paragraph{Case study.} To more meticulously examine the optimization effects of \ours{} on intermediate documents and software artifacts, we selected sliding block puzzle scenario for a detailed case study. In sliding block puzzle games, the user is required to restore a scrambled board to the sequential order of tiles starting from 1 to achieve success. Due to space constraints, we only compared the documents generated from the original prompts with those produced after optimizing the product manager agent (i.e., Experiment 1 and Experiment 2 in \autoref{tab:RQ1SD}). 

\answer{1}{
The experimental results for RQ1 are summarized using two complementary evaluation approaches: LLM-as-a-judge document evaluation and human evaluation of software quality.

In the LLM-as-a-judge evaluation, agent optimization led to consistent improvements in the quality of both PRD and SDD documents. Specifically, SDD consistency increased from 4.4 to 4.7, while SDD completeness and communicability improved from 3.45 to 3.75 and from 4.15 to 4.55, respectively. In addition, PRD completeness increased from 3.85 to 4.25, and clarity improved from 4.05 to 4.45, indicating systematic enhancements in document quality.

In the human evaluation, agent optimization resulted in a clear increase in perceived software quality and user satisfaction. Overall user satisfaction rose from 4.5 to 5.75, accompanied by improvements in interface quality (from 4.75 to 5.83), usability (from 4.96 to 5.42), and information quality (from 4.83 to 5.75).

Taken together, results from both automated and human evaluations consistently demonstrate that agent-level optimization leads to substantial improvements in both software artifacts and users’ perceived software quality, thereby providing a clear affirmative answer to RQ1.
}

\subsection{RQ2: Prompter effectively optimizes user prompts compared
to baseline approaches.}
For RQ2, we employ three baselines for comparison: the Naive prompt, which corresponds to the user prompt for each scenario listed in Table 1; the zeroshot Chain-of-Thought prompt~\cite{wei2022chain}, where we further refine the original prompt using a zero-shot approach to generate more detailed instructions; and MetaGPT, which we utilize as a baseline for requirements elicitation. MetaGPT employs built-in agents to develop requirements within the domain of software engineering. Our experimental results are presented in \autoref{tab:GameRQ2} and \autoref{tab:OfficeRQ2}.

\paragraph{Analysis of Results.}  Our dataset can be divided into two subsets: games and tools. In the games subset, our method significantly outperforms the baseline across all four evaluation metrics. Specifically, it achieve scores of 6.30, 6.17, 6.40, and 6.53 (out of 7) in overall user satisfaction, usability, information quality, and interface quality, respectively. This indicates that \ours{} effectively optimizes prompt words. However, in the tools subset, although our method achieve a high overall user satisfaction score of 6.50 and an information quality score of 6.13, it did not yield the best performance in usability and interface quality. This may be attributed to the lower interactivity and functional complexity of the tools subset, which diminishs the performance gaps between different methods.

\begin{table}[tbp]
\centering
\scriptsize
\caption{Comparison of different methods on Game Scenarios: \ours{} consistently outperforms the baseline across all metrics.}
\setlength{\tabcolsep}{2pt}
\label{tab:GameRQ2}
\begin{adjustbox}{max width=\textwidth}
\begin{tabular}{l|cccc|cccc|cccc|cccc}
\toprule
Application Scenario & \multicolumn{4}{c}{Naive} & \multicolumn{4}{c}{Chain-of-thought} & \multicolumn{4}{c}{MetaGPT} & \multicolumn{4}{c}{Ours} \\
\cmidrule(lr){2-5} \cmidrule(lr){6-9} \cmidrule(lr){10-13} \cmidrule(lr){14-17}
& Ove. & Use. & Info. & Inte. & Ove. & Use. & Info. & Inte. & Ove. & Use. & Info. & Inte. & Ove. & Use. & Info. & Inte. \\
\midrule
go & 7.00 & 5.67 & 6.67 & 6.67 & 6.00 & 5.67 & 6.00 & 6.00 & 6.00 & 5.33 & 6.00 & 6.00 & 7.00 & 5.83 & 6.00 & 6.67 \\
chess & 2.00 & 4.92 & 4.33 & 5.33 & 4.00 & 5.25 & 4.67 & 5.33 & 4.00 & 5.17 & 6.67 & 5.33 & 3.00 & 5.25 & 4.67 & 5.33 \\
2048 & 3.00 & 3.42 & 4.00 & 4.00 & 6.00 & 5.33 & 6.67 & 6.00 & 5.00 & 5.17 & 4.67 & 5.00 & 6.00 & 5.25 & 6.67 & 6.00 \\
tetris & 6.00 & 5.25 & 6.33 & 5.67 & 6.00 & 6.00 & 6.33 & 5.67 & 4.00 & 5.00 & 5.00 & 5.33 & 7.00 & 6.33 & 7.00 & 6.33 \\
blackjack & 4.00 & 5.08 & 5.33 & 5.00 & 6.00 & 5.42 & 6.00 & 6.00 & 5.00 & 5.08 & 6.00 & 5.33 & 6.00 & 5.58 & 6.00 & 6.33 \\
hanoi & 4.00 & 4.75 & 5.00 & 6.00 & 4.00 & 3.92 & 4.67 & 4.67 & 3.00 & 4.25 & 5.00 & 4.00 & 6.00 & 5.92 & 5.67 & 6.67 \\
snake & 2.00 & 2.92 & 1.00 & 3.33 & 6.00 & 6.00 & 6.67 & 6.00 & 3.00 & 4.42 & 6.67 & 4.00 & 7.00 & 7.00 & 7.00 & 7.00 \\
minesweeper & 1.00 & 3.00 & 3.00 & 7.00 & 4.00 & 3.92 & 4.67 & 4.67 & 6.00 & 6.08 & 6.33 & 6.00 & 7.00 & 7.00 & 7.00 & 7.00 \\
tic tac toe & 1.00 & 1.75 & 1.67 & 1.00 & 6.00 & 5.75 & 6.67 & 6.00 & 4.00 & 4.75 & 6.00 & 5.00 & 7.00 & 6.50 & 7.00 & 7.00 \\
sliding block & 1.00 & 1.50 & 1.00 & 3.00 & 6.00 & 5.83 & 6.00 & 6.00 & 5.00 & 4.50 & 4.67 & 5.67 & 7.00 & 7.00 & 7.00 & 7.00 \\
\midrule
Avg & 3.10 & 3.83 & 3.83 & 4.70 & 5.40 & 5.31 & 5.83 & 5.63 & 4.40 & 4.88 & 5.57 & 5.00 & \textbf{6.30} & \textbf{6.17} & \textbf{6.40} & \textbf{6.53} \\
\bottomrule
\end{tabular}
\end{adjustbox}
\end{table}

\begin{table}[tbp]
\centering
\scriptsize
\caption{Comparison of different methods on Office Scenarios: \ours{} yields the highest overall satisfaction.}
\setlength{\tabcolsep}{2pt}
\label{tab:OfficeRQ2}
\begin{adjustbox}{max width=1.1\textwidth}
\begin{tabular}{l|cccc|cccc|cccc|cccc}
\toprule
\textbf{Application Scenario} & \multicolumn{4}{c}{\textbf{Naive}} & \multicolumn{4}{c}{\textbf{Chain-of-thought}} & \multicolumn{4}{c}{\textbf{MetaGPT}} & \multicolumn{4}{c}{\textbf{Ours}} \\
\cmidrule(lr){2-5} \cmidrule(lr){6-9} \cmidrule(lr){10-13} \cmidrule(lr){14-17}
 & Ove. & Use. & Info. & Inte. & Ove. & Use. & Info. & Inte. & Ove. & Use. & Info. & Inte. & Ove. & Use. & Info. & Inte. \\
\midrule
clock in & 5.00 & 4.50 & 6.33 & 4.67 & 6.00 & 5.58 & 5.33 & 6.00 & 6.00 & 5.00 & 6.33 & 6.33 & 6.00 & 4.92 & 6.33 & 5.33 \\
image format conversion & 7.00 & 4.42 & 5.67 & 6.00 & 5.00 & 5.50 & 6.00 & 4.67 & 6.00 & 5.33 & 5.67 & 6.00 & 7.00 & 4.83 & 6.00 & 5.67 \\
calculator & 7.00 & 6.50 & 7.00 & 7.00 & 7.00 & 6.50 & 7.00 & 7.00 & 7.00 & 6.17 & 7.00 & 7.00 & 7.00 & 6.50 & 7.00 & 7.00 \\
recording & 6.00 & 5.83 & 5.33 & 5.67 & 6.00 & 5.75 & 6.67 & 6.00 & 6.00 & 5.50 & 6.67 & 6.00 & 6.00 & 6.17 & 5.67 & 5.67 \\
alarming & 7.00 & 4.58 & 5.67 & 5.00 & 6.00 & 5.08 & 6.33 & 6.33 & 5.00 & 4.92 & 6.33 & 5.00 & 7.00 & 5.00 & 6.00 & 5.00 \\
sentiment analysis & 7.00 & 5.58 & 5.33 & 6.67 & 6.00 & 5.58 & 7.00 & 6.00 & 6.00 & 5.50 & 6.67 & 5.67 & 7.00 & 5.67 & 5.67 & 6.67 \\
charting & 6.00 & 4.25 & 6.00 & 4.33 & 7.00 & 5.67 & 6.00 & 6.67 & 6.00 & 4.08 & 5.67 & 4.67 & 6.00 & 4.92 & 6.00 & 5.33 \\
personal website & 5.00 & 5.17 & 5.67 & 5.00 & 6.00 & 5.08 & 6.33 & 5.67 & 7.00 & 6.50 & 7.00 & 7.00 & 7.00 & 6.17 & 6.67 & 6.67 \\
word cloud & 7.00 & 6.17 & 5.33 & 6.67 & 6.00 & 6.17 & 4.67 & 6.33 & 6.00 & 5.75 & 4.67 & 6.33 & 7.00 & 5.83 & 5.33 & 6.00 \\
drawing & 5.00 & 4.75 & 6.33 & 5.00 & 6.00 & 5.42 & 5.00 & 6.67 & 6.00 & 5.17 & 6.00 & 6.00 & 5.00 & 5.25 & 6.67 & 5.33 \\
\midrule
Avg & 6.20 & 5.18 & 5.87 & 5.60 & 6.10 & \textbf{5.63} & 6.03 & \textbf{6.13} & 6.10 & 5.39 & 6.20 & 6.00 & \textbf{6.50} & 5.53 & \textbf{6.13} & 5.87 \\
\bottomrule
\end{tabular}
\end{adjustbox}
\end{table}
\answer{2}{
\ours{} effectively optimizes user prompts in the field of software engineering compare to baseline approaches. For the gaming subset, it achieve scores of 6.30, 6.17, 6.40, and 6.53 in overall user satisfaction, usability, information quality, and interface quality, respectively, demonstrating its efficacy in user prompt optimization. Lower functional complexity may lead to a narrower performance gap among different solutions.
}

\subsection{RQ3: Absence of any stage leads to a decline in task performance to varying extents.}

\paragraph{Ablation study.} Considering the actual operational costs, we conduct our ablation study only on the document ranking experiments in RQ1. To obtain more observable conclusions, we perform the ablation under the same settings as Experiment 1 in RQ1, i.e., we only optimized the TeamLeader agent. In the ablation study, we conducted four separate experiments for the four components of requirements engineering to validate the effectiveness of each module. Our experimental results are presented in \autoref{tab:PRDRQ3}.

\paragraph{Analysis of Results.} As shown in the table, for both PRD and SDD tasks, the absence of any one of the four stages leads to a relative decline in performance, and the order of the extent of decline is consistent across the scoring tasks for both documents. Specifically, the absence of the validation stage has the smallest impact on the results, followed by the Requirements Specification and Requirements Elicitation stages. In both experiments, the absence of the Requirements Analysis stage result in the largest performance drop, demonstrating the effectiveness of using a software requirements specification draft to articulate user needs. The experiments indicate that Requirements Elicitation and Requirements Analysis are the stages with the most significant impact on the results, further validating the necessity of incorporating requirements engineering into prompt engineering in the field of LLM-based software engineering. Additionally, we observe that although the absence of Requirements validation has a marginal impact on document scoring in the SDD task, it still causes a noticeable performance decline in the PRD task. Furthermore, in the PRD experiment, the absence of any stage leads to a comprehensive decline across all three metrics, completeness, consistency, and clarity. In the SDD experiment, however, the absence of the validation stage primarily results in a decline in consistency scores, indicating that the main role of the Requirements validation stage in the SDD document scoring experiment is to enhance the consistency of the output document. Similarly, it can be inferred that the primary role of Requirements Specification is to improve the integrity of the output document, likely because converting software requirements specifications into prompts encourages the LLM to further supplement relevant knowledge. The main roles of the Requirements Elicitation and Requirements Analysis stages are to enhance the integrity and cohesion of the output document, suggesting that methodologies from requirements engineering can effectively enrich the content of the output document while also structuring it more precisely.

\answer{3}{
In \ours{}, the absence of any stage leads to a decline in task performance, with Requirements Analysis and Requirements Elicitation having the most substantial impact on the outcomes. This underscores the necessity of integrating requirements engineering into LLM-based software engineering. Further analysis reveals that the Requirements Elicitation and Requirements Analysis stages primarily enhance the integrity and cohesion of the output document. The Requirements Specification stage mainly contributes to improving the integrity of the output, while the Requirements validation stage plays a key role in enhancing the consistency of the generated document.
}
\begin{table}[tbp]
\centering
\scriptsize
\setlength{\tabcolsep}{1pt}
\caption{Evaluation results of PRD in ablation study : absence of any stage leads to a decline in task performance to varying extents.}
\label{tab:PRDRQ3}
\renewcommand{\arraystretch}{1.2}
\begin{adjustbox}{max width=\textwidth}
\begin{tabular}{l|ccc|ccc|ccc|ccc|ccc|ccc|ccc|ccc}
\toprule
& \multicolumn{12}{c|}{PRD experiment} & \multicolumn{12}{c}{SDD experiment} \\
\hline
\textbf{Application Scenario} & \multicolumn{3}{c|}{\textbf{No validation}} & \multicolumn{3}{c|}{\textbf{No Specification}} & \multicolumn{3}{c|}{\textbf{No Elicitation}} & \multicolumn{3}{c|}{\textbf{No Analysis}} & \multicolumn{3}{c|}{\textbf{No validation}} & \multicolumn{3}{c|}{\textbf{No Specification}} & \multicolumn{3}{c|}{\textbf{No Elicitation}} & \multicolumn{3}{c}{\textbf{No Analysis}}\\
\cline{2-25}
 & Comp. & Cla. & Coh. & Comp. & Cla. & Coh. & Comp. & Cla. & Coh. & Comp. & Cla. & Coh. & Int. & Comm. & Con. & Int. & Comm. & Con. & Int. & Comm. & Con. & Int. & Comm. & Con. \\
\hline
go & 4 & 5 & 4 & 4 & 5 & 4 & 4 & 4 & 4 & 4 & 4 & 4 & 3 & 4 & 5 & 4 & 4 & 4 & 5 & 4 & 5 & 3 & 5 & 4 \\
chess & 3 & 5 & 4 & 3 & 4 & 4 & 4 & 4 & 4 & 1 & 3 & 1 & 4 & 4 & 4 & 4 & 5 & 4 & 4 & 5 & 5 & 1 & 5 & 3 \\
2048 & 4 & 4 & 4 & 4 & 5 & 4 & 4 & 4 & 4 & 5 & 4 & 4 & 4 & 5 & 4 & 3 & 4 & 4 & 3 & 5 & 4 & 4 & 4 & 4 \\
tetris & 4 & 4 & 4 & 4 & 4 & 4 & 4 & 4 & 4 & 4 & 4 & 4 & 4 & 4 & 4 & 4 & 5 & 5 & 4 & 4 & 4 & 4 & 4 & 4 \\
blackjack & 4 & 4 & 4 & 4 & 4 & 5 & 4 & 4 & 4 & 4 & 4 & 4 & 4 & 4 & 4 & 4 & 4 & 5 & 4 & 4 & 4 & 4 & 5 & 4 \\
hanoi & 4 & 4 & 4 & 4 & 4 & 4 & 4 & 4 & 4 & 4 & 4 & 5 & 4 & 4 & 4 & 4 & 5 & 5 & 4 & 5 & 4 & 4 & 4 & 5 \\
snake & 5 & 4 & 4 & 4 & 4 & 4 & 4 & 4 & 5 & 4 & 4 & 4 & 4 & 4 & 5 & 3 & 4 & 5 & 5 & 4 & 4 & 3 & 4 & 5 \\
minesweeper & 4 & 4 & 4 & 4 & 4 & 4 & 4 & 4 & 4 & 4 & 4 & 4 & 4 & 5 & 4 & 4 & 4 & 5 & 4 & 4 & 4 & 4 & 4 & 4 \\
tic tac toe & 4 & 5 & 5 & 5 & 4 & 5 & 4 & 4 & 4 & 4 & 4 & 4 & 4 & 4 & 5 & 3 & 4 & 4 & 4 & 5 & 4 & 3 & 5 & 4 \\
sliding block & 4 & 5 & 5 & 4 & 5 & 4 & 4 & 4 & 4 & 5 & 5 & 4 & 4 & 5 & 5 & 4 & 4 & 5 & 4 & 4 & 4 & 4 & 4 & 4 \\
clock in & 4 & 5 & 4 & 3 & 4 & 4 & 4 & 4 & 4 & 4 & 5 & 4 & 4 & 5 & 4 & 4 & 5 & 4 & 1 & 5 & 5 & 4 & 4 & 5 \\
image format conversion & 4 & 4 & 5 & 4 & 5 & 4 & 4 & 4 & 4 & 5 & 4 & 4 & 4 & 4 & 5 & 4 & 4 & 5 & 3 & 4 & 5 & 4 & 5 & 5 \\
calculator & 4 & 4 & 4 & 4 & 4 & 4 & 4 & 4 & 5 & 4 & 4 & 4 & 3 & 4 & 5 & 4 & 5 & 4 & 4 & 4 & 4 & 4 & 4 & 4 \\
recording & 4 & 4 & 5 & 4 & 4 & 5 & 4 & 5 & 5 & 4 & 4 & 4 & 4 & 5 & 5 & 4 & 5 & 4 & 4 & 4 & 4 & 3 & 4 & 4 \\
alarming & 4 & 4 & 4 & 4 & 4 & 4 & 4 & 4 & 4 & 4 & 5 & 4 & 4 & 4 & 4 & 3 & 5 & 4 & 4 & 5 & 4 & 4 & 4 & 4 \\
sentiment analysis & 4 & 4 & 4 & 4 & 4 & 4 & 4 & 5 & 4 & 4 & 4 & 5 & 4 & 4 & 4 & 4 & 4 & 5 & 4 & 4 & 4 & 3 & 5 & 4 \\
charting & 4 & 4 & 4 & 4 & 5 & 4 & 4 & 4 & 5 & 4 & 4 & 4 & 4 & 4 & 5 & 3 & 4 & 5 & 4 & 4 & 4 & 4 & 5 & 4 \\
personal website & 4 & 4 & 5 & 4 & 4 & 4 & 4 & 4 & 4 & 4 & 4 & 4 & 4 & 5 & 5 & 4 & 4 & 4 & 4 & 4 & 4 & 4 & 4 & 5 \\
word cloud & 3 & 4 & 5 & 4 & 4 & 5 & 4 & 5 & 4 & 4 & 4 & 4 & 4 & 4 & 4 & 4 & 4 & 5 & 3 & 4 & 5 & 3 & 5 & 4 \\
drawing & 4 & 4 & 4 & 4 & 4 & 4 & 4 & 4 & 4 & 4 & 5 & 4 & 4 & 4 & 4 & 4 & 4 & 4 & 4 & 4 & 4 & 4 & 5 & 4 \\
\hline
\textbf{Average} & 3.95 & 4.25 & 4.3 & 3.95 & 4.25 & 4.2 & 4 & 4.15 & 4.2 & 4 & 4.15 & 3.95 & 3.9 & 4.3 & 4.45 & 3.75 & 4.35 & 4.5 & 3.8 & 4.3 & 4.25 & 3.55 & 4.45 & 4.2 \\
\hline
\end{tabular}
\end{adjustbox}
\end{table}

\subsection{RQ4: Our method consistently enhances quality of SDD and PRD across different foundation LLMs.}
To observe the performance differences of \ours{} across various base LLMs, we conduct extra experiments on GPT-5 and GPT-4. Considering computational and manual costs, comparative experiments are performed only on the Team Leader agent in MetaGPT for the document generation task. The results are presented in \autoref{tab:model_comparison}. 

\paragraph{Reuslts.} Our method consistently leads to overall performance improvements on GPT-5~\cite{OpenAI_GPT5_2025}, GPT-4~\cite{openai2023gpt4}, and Qwen2.5-Max, indicating that it effectively optimizes agent prompts across different base LLMs. On GPT-5, the integrity score and the communicativeness score in SDD scoring tasks increase from 3.95 to 4.1 and increases from 4.25 to 4.4 respectively. On the PRD task, the largest gain is observed in cohesiveness, which rises from 4.15 to 4.3, suggesting that our approach enhances the structural quality and logical consistency of generated content. When GPT-4 is used as the base model, we find that the optimized team leader improves the consistency score on SDD from 4.15 to 4.35 and improves the communicativeness score on SDD from 4.05 to 4.2, while also yielding comprehensive gains in clarity, and cohesiveness on PRD. This pattern implies that optimization improves the overall readability and informativeness of the artifacts produced. Using Qwen2.5-Max as the base model, we observe that on SDD, integrity increases from 3.45 to 3.75 and communicativeness improves from 4.35 to 4.55. In the PRD evaluation, completeness rises from 3.85 to 4.25 and clarity from 4.05 to 4.45, reflecting that our method enhances both factual correctness and the clarity of expression in outputs.

\answer{4}{Our method consistently enhances quanlity of SDD and PRD across GPT-5, GPT-4, and Qwen2.5-Max, demonstrating its robustness and generalizability in optimizing agent prompts for different base LLMs.}

\begin{table}[tbp]
\centering
\caption{How performance of \ours{} differs for different LLM.}
\label{tab:model_comparison}
\begin{adjustbox}{max width=\textwidth}
\begin{tabular}{l|l|ccc|l|ccc}
\hline
\multirow{2}{*}{LLM} & \multicolumn{4}{c|}{SDD Setting} & \multicolumn{4}{c}{PRD Setting} \\
\cline{2-9}

 & Optimized agent-document & Int. & Comm. & Con. & Optimized agent-document & Comp. & Cla. & Coh. \\
\hline
\multirow{2}{*}{GPT-5} 
 & None-SDD & 3.95 & 4.25 & 4.5 & None-PRD & 3.9 & 4.4 & 4.15 \\
 & Team leader-SDD   & 4.1  & 4.4  & 4.5 & Team leader-PRD   & 4.0 & 4.45 & 4.3 \\
\cline{1-9}
\multirow{2}{*}{GPT-4} 
 & None-SDD & 3.4 & 4.05  & 4.15 & None-PRD & 3.85 & 4.15 & 4.1 \\
 & Team leader-SDD   & 3.5 & 4.2 & 4.35 & Team leader-PRD   & 3.85  & 4.3 & 4.2 \\
\cline{1-9}
\multirow{2}{*}{Qwen2.5-Max} 
 & None-SDD & 3.45 & 4.35 & 4.4 & None-PRD & 3.85 & 4.05 & 4.1 \\
 & Team leader-SDD   & 3.75 & 4.55 & 4.35 & Team leader-PRD  & 4.25 & 4.45 & 4.15 \\
\hline
\end{tabular}
\end{adjustbox}
\end{table}

\begin{table}[tbp]
\centering
\renewcommand{\arraystretch}{1.2}
\setlength{\tabcolsep}{2pt}
\caption{Comparison of Architect Agent Prompt Optimization With and Without Template Separation.}
\label{tab:discussion}
\scalebox{0.8}{
\begin{tabular}{c|c|*{20}{w{c}{0.3cm}}|c}
\hline
\textbf{Group} & \textbf{Metric / Task id} & 1 & 2 & 3 & 4 & 5 & 6 & 7 & 8 & 9 & 10 & 11 & 12 & 13 & 14 & 15 & 16 & 17 & 18 & 19 & 20 & avg \\
\hline
\multirow{3}{*}{template in optimization} 
 & Comp. & 3 & 2 & 3 & 3 & 4 & 4 & 4 & 4 & 4 & 4 & 4 & 3 & 3 & 4 & 4 & 4 & 3 & 3 & 3 & 4 & 3.52 \\
 & Cla. & 5 & 4 & 4 & 4 & 4 & 4 & 4 & 4 & 4 & 4 & 4 & 5 & 4 & 5 & 4 & 4 & 4 & 4 & 4 & 4 & 4.15 \\
 & Coh. & 5 & 5 & 4 & 5 & 5 & 4 & 4 & 4 & 4 & 4 & 5 & 5 & 5 & 4 & 5 & 4 & 4 & 5 & 5 & 4 & 4.52 \\
\hline
\multirow{3}{*}{template not in optimization} 
 & Comp. & 3 & 3 & 4 & 3 & 3 & 4 & 4 & 4 & 4 & 3 & 5 & 4 & 4 & 3 & 4 & 3 & 4 & 3 & 4 & 3 & 3.6 \\
 & Cla. & 5 & 4 & 4 & 5 & 4 & 5 & 5 & 5 & 4 & 4 & 5 & 4 & 4 & 4 & 5 & 5 & 4 & 4 & 5 & 4 & 4.45 \\
 & Coh. & 4 & 5 & 4 & 5 & 4 & 5 & 4 & 4 & 4 & 5 & 4 & 4 & 4 & 4 & 4 & 5 & 4 & 5 & 5 & 5 & 4.4 \\
\hline
\end{tabular}
}
\end{table}

\section{Discussion}
\label{discussion}
\subsection{The Impact of Templates on Agent Role  Prompt Optimization}  
To better constrain the outputs of large language models, prompt engineers often employ output templates to regulate the structure and content of the generated responses. These templates, which encapsulate specific details and formatting requirements, represent both a concentration of human expertise and potentially mandatory communication protocols that agents must adhere to in industrial applications. Therefore, such templates should likely be kept separate from the agent optimization process to prevent unintended alterations during refinement. To verify whether this assumption holds, we conduct experiments on the architect agent in metaGPT as original system prompt of architect agent contains tamplates. Specifically, we perform two optimization procedures: one in which the template is included as part of the input during optimization, and another in which the template is kept separate,only the optimized prompt is combined with the original template afterward. The test results are presented in the \autoref{tab:discussion}.

Our findings indicate that optimizing the output template separately leads to improvements in completeness and clarity. This outcome is rather intuitive, as manually crafted templates inherently enhance the Large Language Model’s responses by structuring the output specification, thereby promoting more comprehensive and articulate answers. We also observe that decoupling the template during optimization results in a reduction of cohesion. We speculate that this occurs because the isolated template falls outside the purview of the agent responsible for optimizing the system prompt. Therefore, we recommend that during system prompt optimization, the template should not only be treated as an independently optimizable component, but also be incorporated into the input perspective of the prompt engineering process. 

\paragraph{The Relationship Between Work Context and Agent Optimization.} In our experiments, we observe that the most significant improvements in downstream task performance for agents do not always stem from direct optimization of the agents themselves. In fact, optimizing upstream agents can also yield considerable enhancement, as upstream agents can further improve downstream agents' performance by refining their work context. This suggests that for agent system prompts, particularly in multi-agent systems, optimizing the prompts alone is insufficient. Instead, the context should also be incorporated into the optimization framework.

\subsection{Threats to validity}
\label{future}
\paragraph{Internal Validity.} The primary threat to internal validity stems from the stochastic nature of LLMs, which can be categorized into two scenarios: the possibility that the LLM fails to correctly complete the documentation generation task. To mitigate the first scenario, we verify whether the SEagent has generated the corresponding documentation after each execution. If no documentation is produced, the process is re-executed, with a maximum of three attempts. Another issue pertains to the model’s potential to generate multiple versions of requirement documents. This is addressed during scoring by consistently employing the highest score among all versions to ensure a fair evaluation. Considering computational costs, our validation experiments for RQ2 and RQ3 were limited to a single agent optimization within the SEagent framework, which may also cause variation in actual use.

\paragraph{External Validity.} Potential threats to external validity mainly arise from the subjective variability in manual scoring and the selection of different LLMs. To address these issues, we employ two evaluators to assess each software artifact during manual scoring, and the average score is taken as the final evaluation result. Another concern is that this study primarily focuses on GPT-5, GPT-4, and Qwen2.5-Max; Performance may vary with different LLMs.
\section{Conclusion}
\label{conclusion}
In summary, we propose Prompter, a prompt optimization framework guided by requirements engineering in the field of software engineering. This framework is designed to optimize both the system prompts for agents and the input prompts from users. Our experiments demonstrate that our method effectively improves both types of prompts, outperforming existing baselines. Furthermore, we explore several promising research directions and expectations under this framework. Our work points to potential avenues for future related research.
\bibliographystyle{ACM-Reference-Format}
\bibliography{sample-base}

@String{Computer = "{IEEE} Computer" }

@String{Springer = "Springer-Verlag" }

@inproceedings{hong2024metagpt,
  title={MetaGPT: Meta programming for a multi-agent collaborative framework},
  author={Hong, Sirui and Zhuge, Mingchen and Chen, Jonathan and Zheng, Xiawu and Cheng, Yuheng and Zhang, Ceyao and Wang, Jinlin and Wang, Zili and Yau, Steven Ka Shing and Lin, Zijuan and others},
  year={2024},
  organization={International Conference on Learning Representations, ICLR}
}

@article{jin2025conceptual,
  title={A Conceptual Framework for Requirements Engineering of Pretrained-Model-Enabled Systems},
  author={Jin, Dongming and Jin, Zhi and Li, Linyu and Chen, Xiaohong},
  journal={arXiv preprint arXiv:2507.13095},
  year={2025}
}

@article{ullrich2025requirements,
  title={From Requirements to Code: Understanding Developer Practices in LLM-Assisted Software Engineering},
  author={Ullrich, Jonathan and Koch, Matthias and Vogelsang, Andreas},
  journal={arXiv preprint arXiv:2507.07548},
  year={2025}
}

@article{dong2023pace,
  title={Pace: Improving prompt with actor-critic editing for large language model},
  author={Dong, Yihong and Luo, Kangcheng and Jiang, Xue and Jin, Zhi and Li, Ge},
  journal={arXiv preprint arXiv:2308.10088},
  year={2023}
}

@article{sinha2024survival,
  title={Survival of the Safest: Towards Secure Prompt Optimization through Interleaved Multi-Objective Evolution},
  author={Sinha, Ankita and Cui, Wendi and Das, Kamalika and Zhang, Jiaxin},
  journal={arXiv preprint arXiv:2410.09652},
  year={2024}
}

@inproceedings{van2000requirements,
  title={Requirements engineering in the year 00: A research perspective},
  author={Van Lamsweerde, Axel},
  booktitle={Proceedings of the 22nd international conference on Software engineering},
  pages={5--19},
  year={2000}
}

@inproceedings{gorer2023generating,
  title={Generating requirements elicitation interview scripts with large language models},
  author={G{\"o}rer, Binnur and Aydemir, Fatma Ba{\c{s}}ak},
  booktitle={2023 ieee 31st international requirements engineering conference workshops (rew)},
  pages={44--51},
  year={2023},
  organization={IEEE}
}

@article{quattrocchi2025can,
  title={Can LLMs Generate User Stories and Assess Their Quality?},
  author={Quattrocchi, Giovanni and Pasquale, Liliana and Spoletini, Paola and Baresi, Luciano},
  journal={arXiv preprint arXiv:2507.15157},
  year={2025}
}

@article{jin2025system,
  title={A System Model Generation Benchmark from Natural Language Requirements},
  author={Jin, Dongming and Jin, Zhi and Li, Linyu and Fang, Zheng and Li, Jia and Chen, Xiaohong},
  journal={arXiv preprint arXiv:2508.03215},
  year={2025}
}

@inproceedings{lutze2024generating,
  title={Generating specifications from requirements documents for smart devices using large language models (llms)},
  author={Lutze, Rainer and Waldh{\"o}r, Klemens},
  booktitle={International Conference on Human-Computer Interaction},
  pages={94--108},
  year={2024},
  organization={Springer}
}

@article{januaritaiso,
  title={ISO/IEC/IEEE 29148-2018},
  author={Januarita, Dwi and Prabowo, Wahyu Adi}
}

@article{wang2025supporting,
  title={Supporting Software Formal Verification with Large Language Models: An Experimental Study},
  author={Wang, Weiqi and Farrell, Marie and Cordeiro, Lucas C and Zhao, Liping},
  journal={arXiv preprint arXiv:2507.04857},
  year={2025}
}

@article{gu2024survey,
  title={A survey on llm-as-a-judge},
  author={Gu, Jiawei and Jiang, Xuhui and Shi, Zhichao and Tan, Hexiang and Zhai, Xuehao and Xu, Chengjin and Li, Wei and Shen, Yinghan and Ma, Shengjie and Liu, Honghao and others},
  journal={arXiv preprint arXiv:2411.15594},
  year={2024}
}

@article{liu2023g,
  title={G-eval: NLG evaluation using gpt-4 with better human alignment},
  author={Liu, Yang and Iter, Dan and Xu, Yichong and Wang, Shuohang and Xu, Ruochen and Zhu, Chenguang},
  journal={arXiv preprint arXiv:2303.16634},
  year={2023}
}

@ARTICLE{5981339,
  author={},
  journal={IEEE Std 1016-2009 (Revision of IEEE Std 1016-1998) - Redline}, 
  title={IEEE Standard for Information Technology--Systems Design--Software Design Descriptions - Redline}, 
  year={2009},
  volume={},
  number={},
  pages={1-58},
  doi={}}

@article{cagan2005write,
  title={How To Write a Good PRD},
  author={Cagan, Martin},
  journal={Silicon Valley Product Group, San Francisco},
  year={2005}
}

@article{or1979universal,
  title={Universal Documentation System Handbook Volume I: System Description Program Introduction},
  author={Or Publisher: Range Commanders Council STEWS-SA-R White Sands Missile Range},
  year={1979}
}

@article{lewis1995computer,
  title={Computer system usability questionnaire},
  author={Lewis, James R},
  journal={International Journal of Human-Computer Interaction},
  year={1995}
}

@misc{youware,
      title={YouWare| First AI Coding Community Where Builders Create}, 
      author={YouWare},
      year={2025},
      url={https://www.youware.com/}, 
}

@misc{huang2025knowledgeguidedmultiagentframeworkautomated,
      title={Knowledge-Guided Multi-Agent Framework for Automated Requirements Development: A Vision}, 
      author={Jiangping Huang and Dongming Jin and Weisong Sun and Yang Liu and Zhi Jin},
      year={2025},
      eprint={2506.22656},
      archivePrefix={arXiv},
      primaryClass={cs.SE},
      url={https://arxiv.org/abs/2506.22656}, 
}

@article{wei2022chain,
  title={Chain-of-thought prompting elicits reasoning in large language models},
  author={Wei, Jason and Wang, Xuezhi and Schuurmans, Dale and Bosma, Maarten and Xia, Fei and Chi, Ed and Le, Quoc V and Zhou, Denny and others},
  journal={Advances in neural information processing systems},
  volume={35},
  pages={24824--24837},
  year={2022}
}

@misc{qwen2.5-max,
  author       = {Qwen Team},
  title        = {Qwen2.5-Max: The Most Capable Model in the Qwen Series},
  year         = {2024},
  howpublished = {\url{https://huggingface.co/Qwen/Qwen2.5-Max}},
  note         = {Accessed: 2025-09-06}
}

@misc{googlewhite,
  author = {Lee Boonstra},
  title  = {Prompt Engineering},
  year   = {2025},
  url    = {https://s.baoyu.io/files/2025-01-18-pdf-1-TechAI-Goolge-whitepaper_Prompt\%20Engineering_v4-af36dcc7a49bb7269a58b1c9b89a8ae1.pdf},
  note   = {Accessed: 2025-04-05}
}

@manual{deepseek_api,
  title        = {DeepSeek API Documentation},
  author       = {{DeepSeek}},
  year         = {2025},
  note         = {Accessed: April 2025},
  url          = {https://platform.deepseek.com/api-docs/}
}

@manual{qwen_api,
  title        = {qwen-api},
  author       = {qwen},
  year         = {2025},
  url          = {https://help.aliyun.com/zh/qwen/developer-reference/quick-start}
}

@book{wymore2018model,
  title={Model-based systems engineering},
  author={Wymore, A Wayne},
  year={2018},
  publisher={CRC press}
}

@inproceedings{do2024prompt,
  title={Prompt optimization via adversarial in-context learning},
  author={Do, Xuan Long and Zhao, Yiran and Brown, Hannah and Xie, Yuxi and Zhao, James Xu and Chen, Nancy and Kawaguchi, Kenji and Shieh, Michael and He, Junxian},
  booktitle={Proceedings of the 62nd Annual Meeting of the Association for Computational Linguistics (Volume 1: Long Papers)},
  pages={7308--7327},
  year={2024}
}

@misc{buildfire2025appstatistics,
  title        = {Mobile App Download Statistics \& Usage Statistics (2025)},
  author       = {Lauren},
  year         = {2024},
  month        = {Dec},
  howpublished = {\url{https://buildfire.com/app-statistics/}},
  note         = {Last updated December 31, 2024},
}

@misc{wifitalents2025appindustry,
  title        = {App Industry Statistics},
  author       = {{WifiTalents Team}},
  year         = {2025},
  month        = {Jun},
  howpublished = {\url{https://wifitalents.com/app-industry-statistics/}},
  note         = {Published June 2, 2025},
}

@misc{jin2025iredevknowledgedrivenmultiagentframework,
      title={iReDev: A Knowledge-Driven Multi-Agent Framework for Intelligent Requirements Development}, 
      author={Dongming Jin and Weisong Sun and Jiangping Huang and Peng Liang and Jifeng Xuan and Yang Liu and Zhi Jin},
      year={2025},
      eprint={2507.13081},
      archivePrefix={arXiv},
      primaryClass={cs.SE},
      url={https://arxiv.org/abs/2507.13081}, 
}

@inproceedings{wermelinger2023using,
  title={Using github copilot to solve simple programming problems},
  author={Wermelinger, Michel},
  booktitle={Proceedings of the 54th ACM Technical Symposium on Computer Science Education V. 1},
  pages={172--178},
  year={2023}
}

@article{ray2025review,
  title={A Review on Vibe Coding: Fundamentals, State-of-the-art, Challenges and Future Directions},
  author={Ray, Partha Pratim},
  journal={Authorea Preprints},
  year={2025},
  publisher={Authorea}
}

@article{yang2023instoptima,
  title={Instoptima: Evolutionary multi-objective instruction optimization via large language model-based instruction operators},
  author={Yang, Heng and Li, Ke},
  journal={arXiv preprint arXiv:2310.17630},
  year={2023}
}

@misc{OpenAI_GPT5_2025,
  author       = {OpenAI},
  title        = {GPT-5},
  howpublished = {\url{https://openai.com/gpt-5/}},
  year         = {2025},

}

@misc{prasad2023gripsgradientfreeeditbasedinstruction,
      title={GrIPS: Gradient-free, Edit-based Instruction Search for Prompting Large Language Models}, 
      author={Archiki Prasad and Peter Hase and Xiang Zhou and Mohit Bansal},
      year={2023},
      eprint={2203.07281},
      archivePrefix={arXiv},
      primaryClass={cs.CL},
      url={https://arxiv.org/abs/2203.07281}, 
}

@misc{openai2023gpt4,
  title        = {GPT-4 Technical Report},
  author       = {{OpenAI}},
  year         = {2023},
  howpublished = {\url{https://arxiv.org/abs/2303.08774}},
  note         = {Accessed: 2025-09-12}
}

@misc{openai_chatml_github,
  author       = {{OpenAI}},
  title        = {ChatML: Chat Markup Language documentation in the OpenAI Python SDK (release v0.28.0)},
  year         = {2023},
  publisher    = {GitHub},
  howpublished = {\url{https://github.com/openai/openai-python/blob/release-v0.28.0/chatml.md}},
  note         = {Accessed: 2026-01-14}
}

@misc{anthropic_claude_api_2026,
  author       = {{Anthropic}},
  title        = {Anthropic Claude API Documentation},
  year         = {2026},
  howpublished = {\url{https://platform.claude.com/docs/en/api/overview}},
  note         = {Accessed: 2026-01-14; Includes messages API reference and request/response examples}
}

@misc{deepseek_api_docs_2026,
  author       = {{DeepSeek AI}},
  title        = {DeepSeek API Documentation},
  year         = {2026},
  howpublished = {\url{https://api-docs.deepseek.com/}},
  note         = {Accessed: 2026-01-14; OpenAI-Compatible Chat API reference}
}

@misc{wang2023promptagentstrategicplanninglanguage,
      title={PromptAgent: Strategic Planning with Language Models Enables Expert-level Prompt Optimization}, 
      author={Xinyuan Wang and Chenxi Li and Zhen Wang and Fan Bai and Haotian Luo and Jiayou Zhang and Nebojsa Jojic and Eric P. Xing and Zhiting Hu},
      year={2023},
      eprint={2310.16427},
      archivePrefix={arXiv},
      primaryClass={cs.CL},
      url={https://arxiv.org/abs/2310.16427}, 
}

@article{browne2012survey,
  title={A survey of monte carlo tree search methods},
  author={Browne, Cameron B and Powley, Edward and Whitehouse, Daniel and Lucas, Simon M and Cowling, Peter I and Rohlfshagen, Philipp and Tavener, Stephen and Perez, Diego and Samothrakis, Spyridon and Colton, Simon},
  journal={IEEE Transactions on Computational Intelligence and AI in games},
  volume={4},
  number={1},
  pages={1--43},
  year={2012},
  publisher={IEEE}
}

@article{lin2024prompt,
  title={Prompt optimization with human feedback},
  author={Lin, Xiaoqiang and Dai, Zhongxiang and Verma, Arun and Ng, See-Kiong and Jaillet, Patrick and Low, Bryan Kian Hsiang},
  journal={arXiv preprint arXiv:2405.17346},
  year={2024}
}

@article{xiang2025self,
  title={Self-supervised prompt optimization},
  author={Xiang, Jinyu and Zhang, Jiayi and Yu, Zhaoyang and Teng, Fengwei and Tu, Jinhao and Liang, Xinbing and Hong, Sirui and Wu, Chenglin and Luo, Yuyu},
  journal={arXiv preprint arXiv:2502.06855},
  year={2025}
}

@inproceedings{do2025makes,
  title={What makes a good natural language prompt?},
  author={Do, Xuan Long and Dinh, Duy and Nguyen, Ngoc-Hai and Kawaguchi, Kenji and Chen, Nancy and Joty, Shafiq and Kan, Min-Yen},
  booktitle={Proceedings of the 63rd Annual Meeting of the Association for Computational Linguistics (Volume 1: Long Papers)},
  pages={5835--5873},
  year={2025}
}

@article{santana2025prompting,
  title={Which Prompting Technique Should I Use? An Empirical Investigation of Prompting Techniques for Software Engineering Tasks},
  author={Santana Jr, EG and Benjamin, Gabriel and Araujo, Melissa and Santos, Harrison and Freitas, David and Almeida, Eduardo and Neto, Paulo Anselmo da and Li, Jiawei and Chun, Jina and Ahmed, Iftekhar},
  journal={arXiv preprint arXiv:2506.05614},
  year={2025}
}

@inproceedings{goguen1993techniques,
  title={Techniques for requirements elicitation},
  author={Goguen, Joseph A and Linde, Charlotte},
  booktitle={[1993] Proceedings of the IEEE International Symposium on Requirements Engineering},
  pages={152--164},
  year={1993},
  organization={IEEE}
}

@book{wiegers2013software,
  title={Software requirements},
  author={Wiegers, Karl and Beatty, Joy},
  year={2013},
  publisher={Pearson Education}
}

@misc{best_practice_openai,
  author       = {{openai}},
  title        = {Best practices for prompt engineering with the OpenAI API},
  year         = {2026},
  howpublished = {\url{https://help.openai.com/en/articles/6654000-best-practices-for-prompt-engineering-with-the-openai-api}}
}

@article{wu2022survey,
  title={A survey of human-in-the-loop for machine learning},
  author={Wu, Xingjiao and Xiao, Luwei and Sun, Yixuan and Zhang, Junhang and Ma, Tianlong and He, Liang},
  journal={Future Generation Computer Systems},
  volume={135},
  pages={364--381},
  year={2022},
  publisher={Elsevier}
}

@article{HajesmaeelGohari2022mHealthQuestionnaires,
  title={The most used questionnaires for evaluating satisfaction, usability, acceptance, and quality outcomes of mHealth services},
  author={Hajesmaeel-Gohari, S. and others},
  journal={BMC Medical Informatics and Decision Making},
  year={2022}
}

@article{RussJara2025GuideUsabilityQuestionnairesHealthIT,
  title={A practical guide to usability questionnaires that evaluate health IT},
  author={Russ-Jara, A. L. and others},
  journal={Journal of the American Medical Informatics Association},
  year={2025}
}
\clearpage
\appendix
\subsection{Computer System Usability Questionnaire}

\begin{table}[h]
\centering
\caption{Computer System Usability Questionnaire (CSUQ) Items}
\label{tab:csuq_items}
\begin{tabular}{p{1.2cm} p{11.5cm}}
\toprule
\textbf{Item} & \textbf{Statement} \\
\midrule
Q1  & I believe that I can use this system effectively. \\
Q2  & The system is easy to use. \\
Q3  & I feel confident when using this system. \\
Q4  & I find the system unnecessarily complex to use. (reverse-scored) \\
Q5  & I think that completing my tasks with this system requires too much effort. (reverse-scored) \\
Q6  & The functions provided by the system meet my needs. \\
Q7  & The system’s functions work in a way that is consistent with how I perform my tasks. \\
Q8  & I think there are too many unnecessary functions in the system. \\
Q9  & The information provided by the system helps me complete my tasks. \\
Q10 & The information provided by the system is clear. \\
Q11 & The information provided by the system is sufficiently detailed for completing my tasks. \\
Q12 & The information provided by the system is easy to understand. \\
Q13 & I never feel confused when using this system. \\
Q14 & The interface layout is well organized and easy to use. \\
Q15 & The size of buttons and controls is appropriate and easy to operate. \\
Q16 & Commonly used functions are easy to find. \\
Q17 & The system responds quickly. \\
Q18 & The system design takes users’ needs into consideration. \\
Q19 & Overall, I am satisfied with the usability and functionality of this system. \\
\bottomrule
\end{tabular}
\vspace{0.5em}

\footnotesize
\textit{Note.} All items are rated on a 7-point Likert scale: 
1 = Very dissatisfied, 
2 = Dissatisfied, 
3 = Slightly dissatisfied, 
4 = Neutral, 
5 = Slightly satisfied, 
6 = Satisfied, 
7 = Very satisfied.
\end{table}

\end{document}